\documentclass[12pt]{iopart}

\usepackage{graphicx}% Include figure files
\usepackage{dcolumn}% Align table columns on decimal point
\usepackage{bm}% bold math
\usepackage{caption}
\usepackage{subfloat}
\usepackage{subfig}
\usepackage{float}
\usepackage{color}

\begin{document}

\title{Phase Transition and Field Effect Topological Quantum Transistor made of monolayer MoS${}_{2}$}


\author{H. Simchi}
\ead{simchi@alumni.iust.ac.ir}
\address{Department of Physics, Iran University of Science and Technology, Narmak, Tehran 16844, Iran}
\address{Semiconductor Technology Center, P.O. Box 19575-199, Tehran, Iran} 

\author{M. Simchi}
\address{Electrical Engineering Department, Sharif University of Technology, Tehran, Iran}

\author{M. Fardmanesh}
\address{Electrical Engineering Department, Sharif University of Technology, Tehran, Iran}

\author{F. M. Peeters}
\ead{francois.peeters@uantwerpen.be}
\address{Departement Fysica, Universiteit Antwerpen, Groenenborgerlaan 171, B-2020 Antwerpen, Belgium}


\date{\today}

\begin{abstract}
We study topological phase transitions and topological quantum field effect transistor in monolayer Molybdenum Disulfide (MoS${}_{2}$) using a two-band Hamiltonian model. Without considering the quadratic ($q^2$) diagonal term in the Hamiltonian, we show that the phase diagram includes quantum anomalous Hall (QAH), quantum spin Hall (QSH), and spin quantum anomalous Hall effect (SQAH) regions such that the topological Kirchhoff law is satisfied in the plane. By considering the $q^2$ diagonal term and including one valley, it is shown that MoS${}_{2}$ has a non-trivial topology, and the valley Chern number is non-zero for each spin. We show that the wave function is (is not) localized at the edges when the $q^2$ diagonal term is added (deleted) to (from) the spin-valley Dirac mass equation. We calculate the quantum conductance of zigzag MoS${}_{2}$ nanoribbons by using the nonequilibrium Green function method and show how this device works as a field effect topological quantum transistor. 
\end{abstract}

\vspace{2pc}
\noindent{\it Keywords}: Phase Transition; MoS${}_{2}$; Topological Insulator; Field Effect Transistor
%\keywords{Phase Transition; MoS${}_{2}$; Topological Insulator; Field Effect Transistor}

\maketitle

\section{Introduction}
Phase transitions are an important tool in the armory of a material scientist. In the simplest sense, a phase diagram demarcates regions of existence of various phases. In addition, a phase can be defined as a physically distinct and chemically homogeneous portion of a system that has a particular chemical composition and structure. As an example, water in liquid or vapor state is single phase, and ice floating on water is an example of a two phase system. 

In 1879, the American physicist E. H. Hall observed the deflected motion of charged particles in solids under external electric and magnetic field \cite{R1}. The effect is called the Hall Effect (HE). In some materials, the electron orbital motion is coupled to its spin, and consequently, a spin-orbit or spin transverse force can be used to understand the spin-dependent scattering by either impurities (extrinsic origin) or band structure (intrinsic origin). Thus, the anomalous Hall Effect (AHE) can have either an extrinsic or an intrinsic origin due to the spin-dependent band structure of the conduction electrons, which can be expressed in terms of the Berry phase in momentum space \cite{R2}. While the AHE vanishes in the absence of an external magnetic field and in the absence of magnetization in a paramagnetic metal, the spin-dependent deflected motion of electrons in a solid can still lead to an observable effect, that is, the spin Hall effect (SHE) \cite{R3, R4}. The quantum Hall effect (QHE) is a quantum version of the Hall Effect in two dimensions. The key feature of the QHE is that all electrons in the bulk are localized and the electrons near the edges form a series of edge-conducting channels \cite{R5}. The quantized anomalous Hall Effect (QAHE) can be realized in a ferromagnetic insulator with strong spin-orbit coupling \cite{R6,R7,R8,R9}. Finally, similar to SHE, the spin version of QHE is called the quantum spin Hall effect (QSHE). The QSHE can be regarded as a combination of two quantum anomalous Hall effects of spin-up and spin-down electrons with opposite chirality \cite{Rten,R11}. 

Here, we want to address the following question: Is it possible that a specific type of HE is changed to another type of HE by applying some specific fields? This implies that a phase transition occurs between different phases of HE. Physicists have already developed the necessary concepts and theories which are necessary for explaining and estimating the phase transition between different phases of HE \cite{R12,R13,R14,R15,R16}. One of the main goals of this article is to study the topological phase transition in monolayer Molybdenum Disulfide (MoS${}_{2}$) and to show how this can be used in a field effect transistor (FET). 

The flow of carriers between source and drain is adjusted by applying an external gate voltage. If the conductance of the channel is quantized, the current will be quantized and the transistor is called a quantum FET (QFET) \cite{R17}. If the quantized conductance is topologically protected, it will be robust against impurities due to its topological stability. Consequently, we can call it a field effect topological quantum transistor (FETQT) \cite{R17}. Basically, one is able to design a three-digit quantum transistor by attaching an antiferromagnet based on such a topological phase transition \cite{R17}. Therefore, another main goal of this article is to investigate the FETQT in a monolayer of MoS${}_{2}$.

The monolayer $MoS_{2}$ has a honeycomb lattice and is a direct band gap semiconductor with band gap $E_{g}=1.86 eV$. The two planes of sulfur atoms are placed above and below the plane of $Mo$ atom.  The $Mo-S$ atoms form an almost ideal trigonal prism structure with a $Mo-Mo$ distance $a=3.16$ \AA and a slight elongation along the perpendicular axis with $Mo-S$ distance, $b=1.586$ \AA\cite{R36}.
\\Different Hamiltonian models have been introduced to describe the electronic properties of MoS${}_{2}$. Usually, the three $4d$-orbitals of the Mo i.e., $d_{{3z}^2-r^2}$ , $d_{xy}$, and $d_{x^2-y^2}$ and three $3p$-orbitals of S i.e., $p_x,p_{y,}$ and $p_z$ and/or their hybridization have been considered and the Slater-Koster \cite{R18} method has been used to find the tight binding Hamiltonian model of MoS${}_{2}$ [19-22]. It has been shown that a low-energy two-band Hamiltonian model can be deduced around the K-points for each spin component \cite{R20,R21,R22}. By considering $|{{\varphi }_c>=d}_{{3z}^2-r^2}$ and $|{\varphi }^{\tau }_v>=d_{x^2-y^2}+i\tau d_{xy}$ as the basis wave vectors, where the subscript $c\left(v\right)\ $indicates conduction (valence) band, and $\tau =\pm 1$ is the valley index, a two-band $\overrightarrow{k}.\overrightarrow{p}\ $Hamiltonian has been introduced \cite{R23,R24,R25,R26}. \\Using first-principles calculations within density functional theory, the intrinsic spin Hall effect in monolayers of group-VI transition-metal dichalcogenides (TMD) MX${}_{2}$ (M = Mo, W and X = S, Se) has been investigated \cite{R27}. It was shown that because of the inversion symmetry breaking and the strong spin-orbit coupling charge carriers in opposite valleys carry opposite Berry curvature and spin moment, giving rise to both a valley-Hall and a spin-Hall effect \cite{R27}. Qian et al showed that the quantum spin Hall phase can be transformed into a trivial one by applying a vertical electric field in $1T^\prime$ structure of TMD \cite{R28}. A weak topological protection for the metallic edge modes in the zigzag MoS${}_{2}$ nanoribbon has been clarified by considering a low-energy $\overrightarrow{k}.\overrightarrow{p}\ $Hamiltonian \cite{R22}. In addition, it has been shown that the crossing point of the edge modes is not located on the K-point and it shifts away from it due to the effect of trigonal warping \cite{R22}. Some of us have shown that, by applying a transverse electric field beyond a critical value, the inverted band gap disappears and the zigzag MoS${}_{2}$ nanoribbon (in $1H$ structure) turns into a semiconductor \cite{R29}. Olsen has applied first principles calculations to show that the quantum spin Hall insulator $1T^\prime$-MoS${}_{2}$ exhibits a phase transition to a trivial insulator upon adsorption of various atoms \cite{R30}. Liu et al used quantum transport device simulations to investigate the potential of single-layer MoS${}_{2}$ FETs for vertical field modulation of the topological edge states \cite{R31}. Experimental and theoretical works have unambiguously confirmed that the contribution of edge states to the channel conductance is significant before the threshold voltage but negligible once the bulk of the TMD device becomes conductive \cite{R32}. 

In this paper, we study the topological phase transition in monolayer MoS${}_{2}$ by using a low-energy two-band Hamiltonian model around the K-points. First, we neglect the electron-hole asymmetry and the $q^2$ diagonal terms in the spin-valley Dirac mass equation and find the phase diagram in the $(V, \Delta M)$  plane where $V$ and $\Delta M$ are external applied voltage and exchange field to the A(B)-sublattice, respectively. We will show that the phase diagram includes QAH, QSH, and SQAH regions such that the topological Kirchhoff law \cite{R15} is satisfied in the plane. Furthermore, we find that the wave function along the width of a zigzag nanoribbon of MoS${}_{2}$ (e.g in the x-direction) is not localized at the edges when $V\left(x\right)=\alpha x\ $ and $\Delta M=0$. By solving the spin-valley Dirac mass equation in case the quadratic ($q^2$) diagonal term is added, it is shown that the wave function becomes localized at the edges. Also, by considering the $q^2$ diagonal term and one valley, it is shown that MoS${}_{2}$ has non-trivial topology and the valley Chern number is non-zero for each spin. Finally, we find the quantum conductance of a zigzag MoS${}_{2}$ nanoribbon by using the nonequilibrium Green function method \cite{R33} and show how this device works as a field effect topological quantum transistor.

The structure of the article is as follows: In section II, the model is introduced. The phase diagram is extracted and the localization of wave functions is studied, in section III. In section IV, the topological quantum transistor behavior of a zigzag nanoribbon is explained and finally the conclusion are presented in section V.

\section{Formulation of the model}
The low-energy two-band Hamiltonian model around the K-points can be written as \cite{R20,R21,R22}:
\begin{equation}
\eqalign{H_{\tau s}\left(q\right)=\frac{{\Delta }_0+{\lambda }_0\tau s}{2}+\frac{\mathrm{\Delta }+\lambda \tau s}{2}{\sigma }_z+t_0a_0\overrightarrow{q}.\ {\sigma }_{\tau }\nonumber\\
+\frac{{\hbar }^2{\left|\overrightarrow{q}\right|}^2}{4m_0}\left(\alpha +\beta {\sigma }_z\right)+{a_0}^2{\left|\overrightarrow{q}\right|}^2({\lambda }^\prime_0+{\lambda }^\prime)\tau s.}
\end{equation}
The two- band Hamiltonian is obtained from a six- band Hamiltonian\cite{R22}. Here, $\Delta$ and $\Delta_{0}$ are crystal fields (energy gap), and $\lambda$, $\lambda_{0}$, $\lambda^{\prime}$, and $\lambda^{\prime}_{0}$ are spin-orbit coupling constants. $\alpha$ is the mass asymmetry parameter and $\beta$ is the topological term and both are related to the general physical properties of the band structure. Finally, $t_{0}$ is the hopping integral and $a_{0}=a/\sqrt{3}$ where $a$ is the lattice constant.
 $s=\pm $ and $\tau =\pm $ stand for the spin and valley degree of freedom, respectively. Notice ${\sigma }_{\tau }=(\tau {\sigma }_x,{\sigma }_y)$ with ${\sigma }_{i=x,y,z}$  are Pauli matrices, $\overrightarrow{q}=(q_x,q_y)$ is the wave vector in two dimensions and  $m_0$ is the free electron mass. The other constants are ${\Delta_{0}=-0.11}$ eV, ${\Delta=1.82}$ eV, ${\lambda_{0}=69}$ meV, ${\lambda=-80}$ meV, ${\lambda_{0}^{\prime}=-17}$ meV, ${\lambda^{\prime}=-2}$ meV, ${t_{0}=2.34}$ eV, ${\alpha=-0.01}$, ${\beta=-1.54}$, and ${\alpha^{\prime}=0.44}$, ${\beta^{\prime}=-0.53}$ which are obtained from Ref. \cite{R22}.

It should be noted that the six and two-band Hamiltonian have been used, before in Refs. \cite {R22,R29,R35}. Some of us have used the six-band Hamiltonian and studied spin-selective transport in a zigzag monolayer ribbon of $MoS_{2}$ using the non-equilibrium Green function (NEGF) method \cite{R29}. Also, they showed that the metallic phase is transferred to a semiconductor phase by applying some external fields \cite {R29}. The two-band Hamiltonian was used to study prefect valley polarization in $MoS_{2}$ using NGEF method in Ref. \cite{R35}. The edge modes in monolayer $MoS_{2}$ have been studied by calculating the normalized projected density of states (NPDOS)\cite{R22} within a two-band Hamiltonian and tight-binding method. The Berry curvature in the whole Brillouin zone was also determined together with Chern number and the time reversal $Z_{2}$ invariant calculated within the $\vec{k}.\vec{p}$ model \cite{R22}. However, they did not study the phase diagram (PD),  the Kirchhoff law in the plane of PD, the field effect topological quantum transistor in monolayer $MoS_{2}$ and they did not find the zero energy wave function under different external conditions. In this article, the above listed quantities will be calculated.

Eq. (1) can be written as
\begin{equation}
H_{\tau s}\left(q\right)=\epsilon \left(k\right)I+\overrightarrow{\sigma }.\overrightarrow{d}  ,                                                                                    
\end{equation} 
where $I$ is the unit matrix,  
\begin{equation}
\epsilon \left(k\right)=\ \frac{{\Delta }_0+{\lambda }_0\tau s}{2}+\frac{{\hbar }^2{\left|\overrightarrow{q}\right|}^2\alpha }{4m_0}+{a_0}^2{\left|\overrightarrow{q}\right|}^2{\lambda }^\prime_0\tau s , 
\end{equation} 
and
\begin{equation}
\eqalign{\overrightarrow{d}={(t}_0a_0\tau q_x)\hat{i}+{(t}_0a_0\tau q_{y)}\hat{j}+(\ \frac{\Delta +\lambda \tau s}{2}+\frac{{\hbar }^2{\left|\overrightarrow{q}\right|}^2\beta }{4m_0}+{a_0}^2{\left|\overrightarrow{q}\right|}^2{\lambda }^\prime\tau s\widehat{)k} .}
\end{equation} 
The Chern number, $C^{\tau }_s$, is defined as \cite{R15}:
\begin{equation}
C^{\tau }_s=\frac{1}{4\pi }\int{d^2q}\left(\frac{\partial \hat{d}}{\partial q_x}\times \frac{\partial \hat{d}}{\partial q_y}\right)\ .\ \widehat{d} . 
\end{equation} 
By using Eq. (4), and neglecting the electron-hole asymmetry, it can be shown:
\begin{equation} 
\left(\frac{\partial \hat{d}}{\partial q_x}\times \frac{\partial \hat{d}}{\partial q_y}\right)\ .\ \hat{d}=\frac{1}{{\left|\overrightarrow{d}\right|}^3}\ t^2_0a^2_0(\ \ \frac{\Delta +\lambda \tau s}{2}-\frac{{\hbar }^2{\left|\overrightarrow{q}\right|}^2\beta }{4m_0}) .                                                    
\end{equation} 
Therefore,
\begin{equation} 
C^{\tau }_s=\frac{1}{4\pi }\int^{\infty }_0{\int^{\infty }_0{\frac{1}{{\left|\overrightarrow{d}\right|}^3}\ t^2_0a^2_0\left(\ \ \frac{\Delta +\lambda \tau s}{2}-\frac{{\hbar }^2{\left|\overrightarrow{q}\right|}^2\beta }{4m_0}\right)dq_xdq_y}} .
\end{equation} 
When solving Eq.(7), we use polar coordinates and write:
\begin{equation}
Cos\theta =\frac{2m_0\left(\Delta +\lambda \tau s\right)+{\hbar }^2{\left|\overrightarrow{q}\right|}^2\beta}{4m_0{\left(t^2_0a^2_0q^2+{\left(\ \frac{\Delta +\lambda \tau s}{2}+\frac{{\hbar}^2{\left|\overrightarrow{q}\right|}^2\beta }{4m_0}\right)}^2\right)}^{1/2}}\ ,
\end{equation} 
then
\begin{equation}
\frac{\partial (Cos\theta)}{\partial({q}^2})= (\frac{-{t}_{0}^2{a}_{0}^2}{2}) \frac{\frac{\Delta+\lambda \tau s}{2}-\frac{\hbar^2 \beta q^2}{4m_{0}}}{d^{3}}
\end{equation}
and therefore
\begin{equation} 
C^{\tau }_s=\frac{1}{4\pi }\int^{\infty }_0\partial(q^2)(\frac{-2}{t_{0}^{2}a_{0}^{2}})\frac{\partial (Cos\theta)}{\partial({q}^2}).
\end{equation}
Now by mapping the Brilliouin zone to the surface of a sphere with radius $\hat{d}$, it can be shown \cite{R22}
\begin{equation} 
C^{\tau }_s=\frac{\tau s}{2}(sgn\left(\Delta +\lambda s\tau \right)-sgn\left(\beta \right)).
\end{equation} 
It means that the Chern number is specified up to the sign of the total spin-valley Dirac mass. Therefore, we should find their signs when we want to plot the phase diagram. Another important subject is the behavior of the position-dependent wave function. We consider a zigzag nanoribbon of MoS${}_{2}$ with its width in the x-direction and its length in the y-direction. If$\ \ \ C=\xi {\sigma }_y$, such that ${\xi }^2=1$, it can be shown that,   ${C\ H}_{\tau s}\ C^{\dagger }=-H^T_{\tau s}$ if 
$\left( \begin{array}{c}
-i\xi {\psi }_B \\ 
i\xi {\psi }_A \end{array}
\right)=\left( \begin{array}{c}
{\psi }_A \\ 
{\psi }_B \end{array}
\right)$ . In other words, the system has particle-hole symmetry (PHS). By using Eq. (1) and  ${\ H}_{\tau s}\left( \begin{array}{c}
{\psi }_A \\ 
{\psi }_B \end{array}
\right)=E\left( \begin{array}{c}
{\psi }_A \\ 
{\psi }_B \end{array}
\right)$, it can be shown (neglecting electron-hole asymmetry) that
\begin{equation} 
\left(\frac{\Delta +\lambda \tau s}{2}+\frac{{\hbar}^2{\left|\overrightarrow{q}\right|}^2\beta }{4m_0}\right){\psi }_A+t_0a_0\left(\tau q_x-iq_y\right){\psi }_B=E{\psi }_A. 
\end{equation} 
Assuming ${\psi }_A=e^{ik_y}{\phi }_A(x)$, substituting into Eq. (12), and using PH-symmetry we find
\begin{equation}
\eqalign{(\frac{\Delta +\lambda \tau s}{2}-\frac{{\hbar}^2\beta }{4m_0}{\partial }^2_x+t_0a_0\tau \xi {\partial }_x){\phi }_{A}\left(x\right)\nonumber \\=(E-\frac{{\hbar}^2k^2_y\beta }{4m_0}-t_0a_0k_y\xi ){\phi }_{A}\left(x\right). }
\end{equation} 
Now for $E=\frac{{\hbar}^2k^2_y\beta }{4m_0}+t_0a_0k_y\xi $, the right hand side of the Eq. (13) is equal to zero, and we obtain
\begin{equation}
\left(\frac{{\hbar}^2\beta }{4m_0}{\partial }^2_x-(\frac{\Delta +\lambda \tau s}{2})-t_0a_0\tau \xi {\partial }_x\right){\phi }_{A}\left(x\right)=0 
\end{equation} 
By solving Eq. (14), we find the position-dependent wave function. In next section, we consider two seperate cases for the electron-hole asymmetry ($\beta$): first, $\beta =0$ and second when $\beta \neq 0$.

\section{Phase diagram}
\subsection{The case $\beta=0$}
We neglect the electron-hole asymmetry term and assume $\beta=0$.  In this case, we can use the $\overrightarrow{k}.\overrightarrow{p}$ Hamiltonian model \cite{R35} and write the total spin-valley Dirac mass equation as \cite{R35}
\begin{equation}
{\mathrm{\Delta }}^{\tau }_s=\frac{\Delta +\lambda \tau s}{2}-V+s\mathrm{\Delta }M , 
\end{equation} 
where, $V$ and $\mathrm{\Delta }M\ $are applied voltage and exchanged field to the A(B)-sublattice, respectively. Fig. 1(a) (Fig. 1(b)) shows the contour plot of ${\mathrm{\Delta }}^{\tau }_s$ for spin up (spin down) in the ($V$, $\mathrm{\Delta }M$) plane with $\Delta =1.82\ $eV and $\lambda =-80$ meV \cite{R22} for the $K$ and $K^\prime$ points. The main difference between $K$ and $K^\prime$ is in the intersection points of lines with the axices.  
Using Eq. (11) and these figures we can find different Chern numbers i.e., C, C${}_{s}$, C${}_{v}$, and C${}_{sv}$ which are Chern, spin-Chern, valley-Chern and spin-valley-Chern numbers, respectively, which are defined as \cite{R12,R13,R14,R15,R16,R17}:
\numparts
\begin{eqnarray}
C=\left(C^K_{\uparrow }+C^K_{\downarrow }\right)+\left(C^{K^\prime}_{\uparrow }+C^{K^\prime}_{\downarrow }\right) ,  \\
2C_s=\left(C^K_{\uparrow }-C^K_{\downarrow }\right)+\left(C^{K^\prime}_{\uparrow }-C^{K^\prime}_{\downarrow }\right) ,   \\
C_v=\left(C^K_{\uparrow }+C^K_{\downarrow }\right)-\left(C^{K^\prime}_{\uparrow }+C^{K^\prime}_{\downarrow }\right) ,   \\
C_{sv}=\left(C^K_{\uparrow }-C^K_{\downarrow }\right)-\left(C^{K^\prime}_{\uparrow }-C^{K^\prime}_{\downarrow }\right) .                                                             
\end{eqnarray}
\endnumparts
\begin{figure}
\captionsetup{singlelinecheck = false, justification=raggedright}
{
    \includegraphics[width=.5\columnwidth]{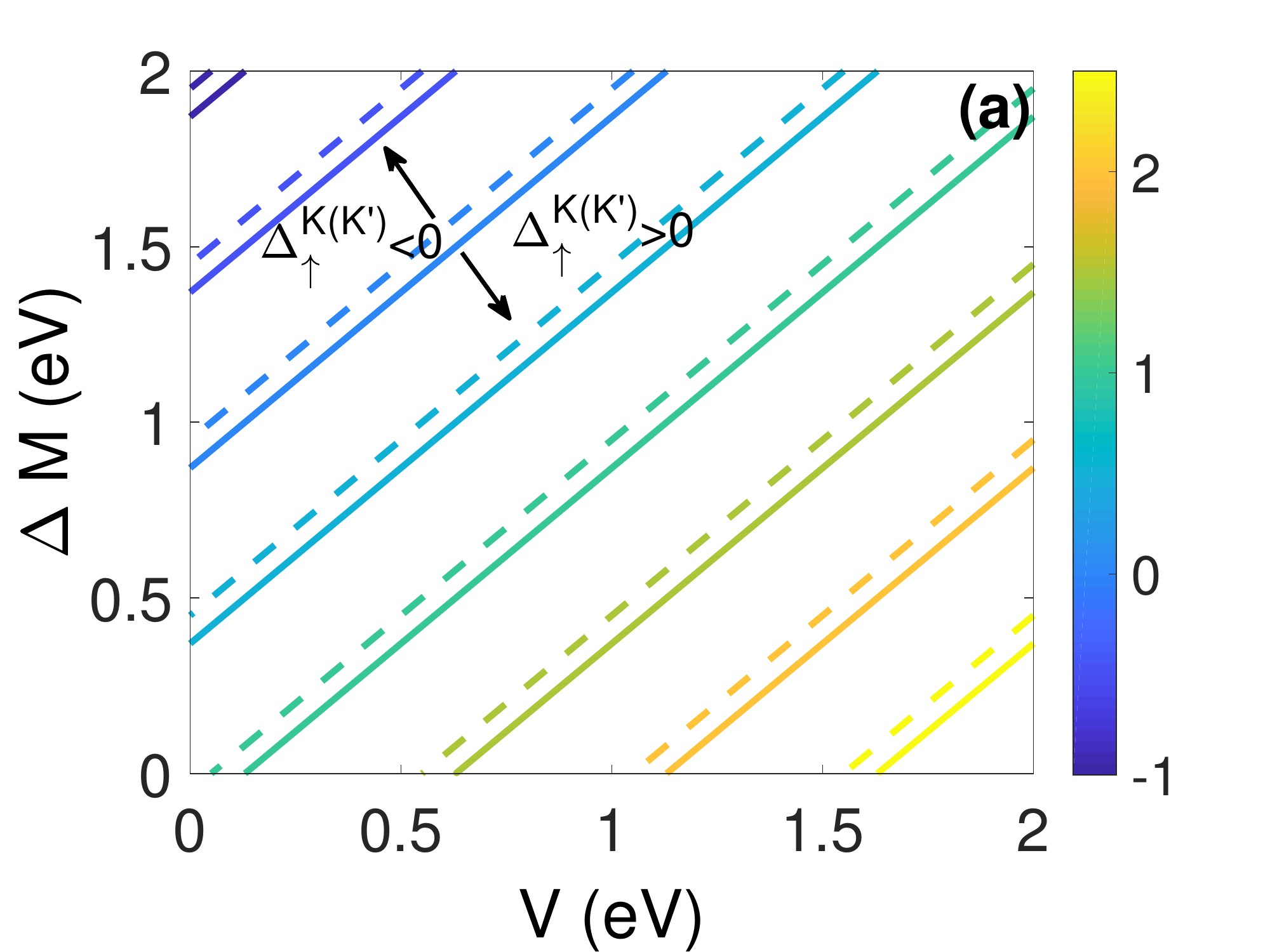}
}
{
    \includegraphics[width=.5\columnwidth]{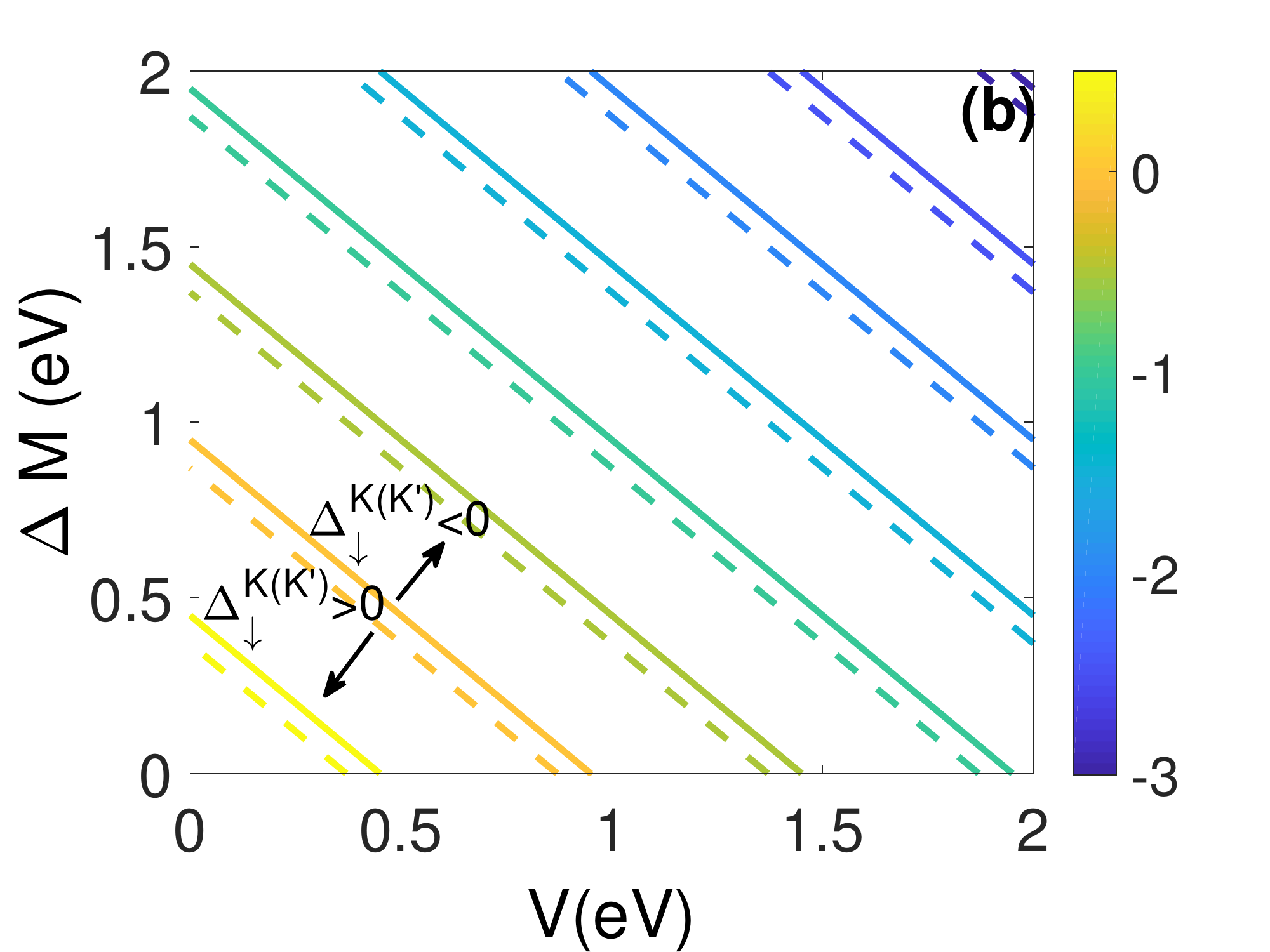}
}
\caption{(Color online) Contour plot of spin-valley Dirac mass term ($\Delta_{s}^{\tau}$) in the plane of applied voltage and exchanged field at $K$ (solid lines) and $K^\prime$ (dashed lines) points for (a) spin up and (b) spin down. Here, $\Delta=1.82\ $ eV and $\lambda =-0.08$ \cite{R22}.}
\end{figure} \\
The results are summarized in Table 1. 
\begin{table}
\captionsetup{singlelinecheck = false, justification=raggedright}
\caption{Chern (C), spin-Chern (C${}_{s}$), Valley-Chern (C${}_{v}$) and spin-valley-Chern numbers (C${}_{sv}$), at the different region of the phase diagram (indicated by Roman letters) (Fig. 2) for $\mathrm{\ }\beta \mathrm{=}0$.}

\begin{tabular}{cccccc} 
\textbf{Region} & \textbf{C} & \textbf{C${}_{s}$} & \textbf{C${}_{v}$} & \textbf{C${}_{sv}$} & \textbf{Type} \\ \hline 
\textbf{I} & 2 & 0 & 0 & 0 & QAH \\
\textbf{II} & 1 & 1/2 & 1 & -1 & SQAH \\ 
\textbf{III} & 0 & 1 & 0 & 0 & QSH \\ 
\textbf{IV} & 1 & 1/2 & 1 & -1 & SQAH \\
\textbf{V} & 0 & 1 & 0 & 0 & QSH \\
\textbf{VI} & 0 & 1 & 0 & 0 & QSH \\ 
\textbf{VII} & 0 & 1 & 0 & 0 & QSH \\  
\textbf{VIII} & -1 & 1/2 & 1 & 1 & SQAH \\ 
\textbf{IX} & -2 & 0 & 0 & 0 & QAH \\
\end{tabular}
\end{table}
Fig. 2 shows the result of Table 1 graphically. Notice that, we can assign four numbers (C, C${}_{s}$, C${}_{v}$, and C${}_{sv}$) to each region of Fig. 2. 
\begin{figure}
\captionsetup{singlelinecheck = false, justification=raggedright}
{
    \includegraphics[width=.5\columnwidth]{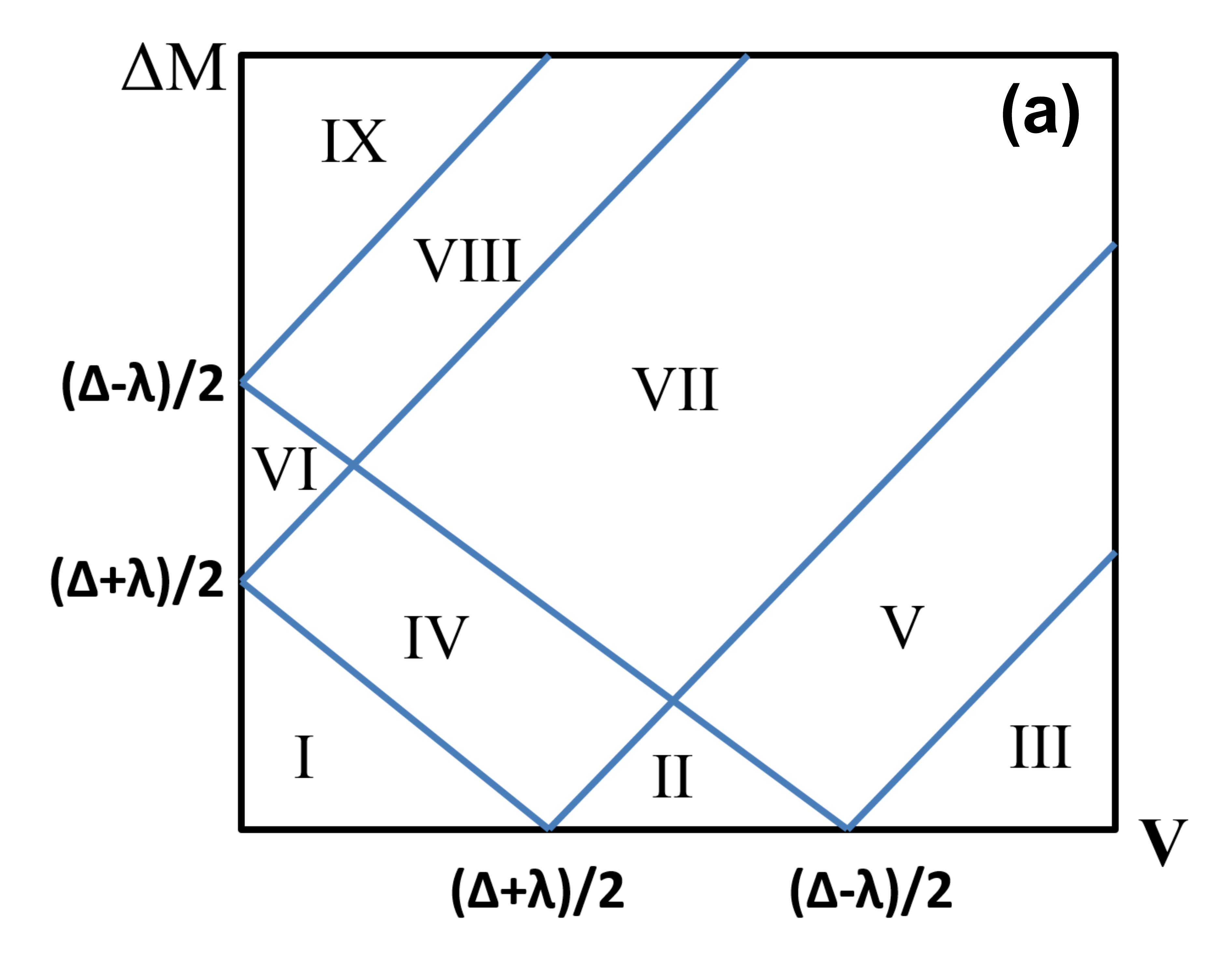}
}
{
    \includegraphics[width=.5\columnwidth]{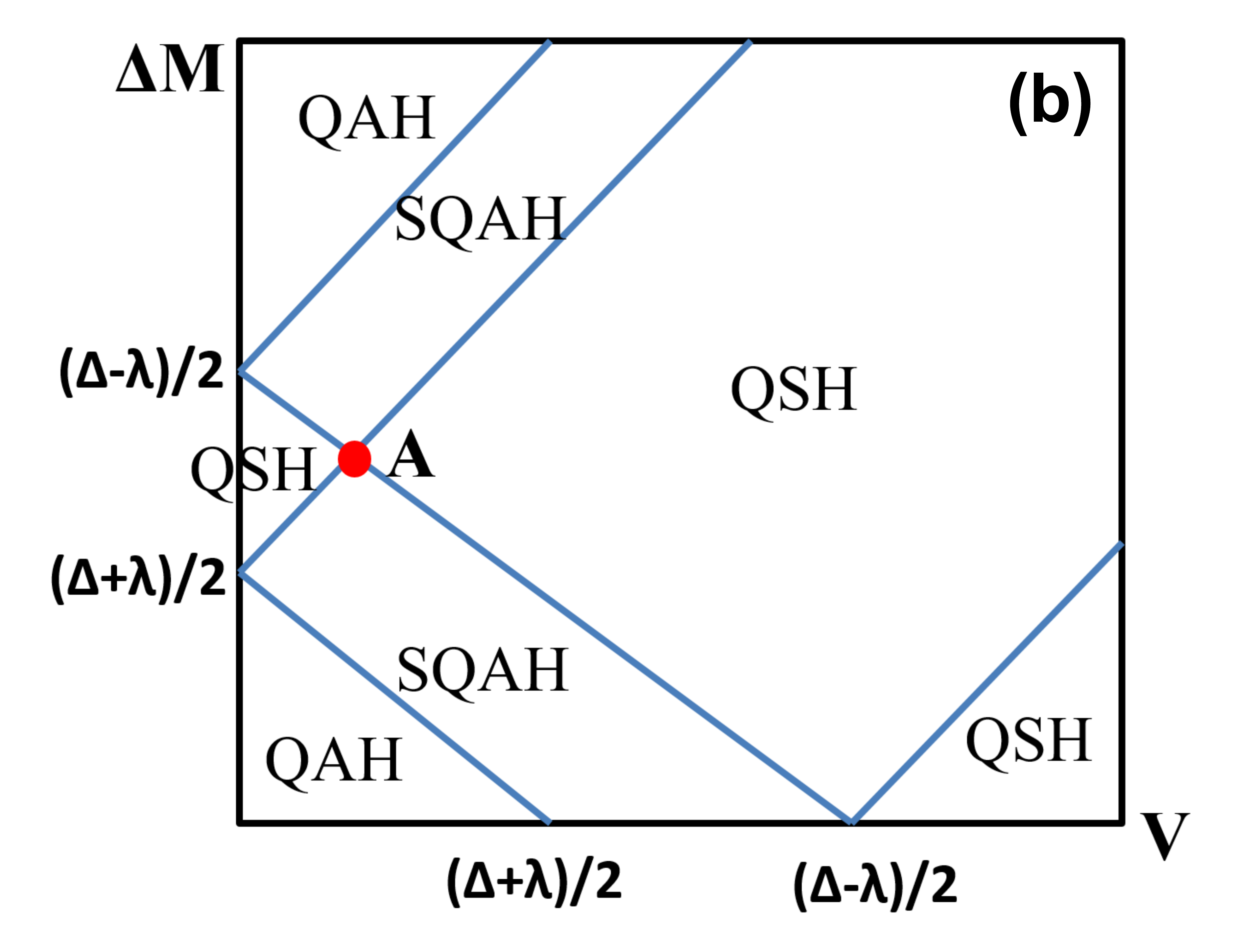}
}
\caption{Phase diagram of MoS${}_{2}$, where we have (a) the different regions and (b) the different types of Hall effects, based on Table 1. Here, $\Delta=1.82\ $ eV and $\lambda =-0.08$ \cite{R22}.}
\end{figure}
\\There are four regions around the point A in Fig. 2(b). Two of theme are QSH and others are SQAH. For justifying the topological Kirchhoff law we should round the point A and subtract the Chern numbers of different regions. Thus, there are two below differences:
\begin{equation} 
{(1,1/2,1,-1)}^{SQAH}-{\left(0,1,0,0\right)}^{QSH}=\left(1,-\frac{1}{2},1,-1\right) , 
\end{equation}
\begin{equation}
{(0,1,0,0)}^{QSH}-{\left(\frac{1,1}{2},\ 1,-1\right)}^{SQAH}=\left(-1,\frac{1}{2},-1,1\right).  
\end{equation} 
Therefore, the topological Kirchhoff law is satisfied at point A (similar to the published results about silicene in Ref. \cite{R15}).
In this case, $E=t_0a_0k_y\xi $ and the general solution of Eq. (14) can be written as ${\phi }_{A\ }\left(x\right)=Be^{f(x)}$ where B is a constant and 
\begin{equation}
f\left(x\right)=-\frac{\tau \xi }{t_0a_0}\int{\frac{\Delta \left(x^\prime\right)+\lambda \tau s}{2}dx^\prime}. 
\end{equation} 
The sign of $\xi $ is determined such that the wave function is finite in the limit $|x|\to \infty $. If the integrand is generally written as
\begin{equation}
\Delta \left(x^\prime\right)=\frac{\Delta +\lambda \tau s}{2}-\ V\left(x^\prime\right), 
\end{equation} 
and $\ V\left(x^\prime\right)\propto \ x^\prime$, then the wave function is localized along the lines $x=\frac{\Delta +\lambda \tau s}{2}=\frac{\Delta }{2}\pm \frac{\lambda }{2}$  and not at the edges of the nanoribbon \cite{R15}.
\subsection{The case $\beta \neq 0$}
From Eq. (11) it is clear that we should specify the sign of $\beta $ if we want to specify the Chern number. In order to obtain the sign of $\beta $, we should study the behavior of the wave function. 
If $\Delta \left(x\right)=a-\ bx$ where $a=\frac{\Delta +\lambda \tau s}{2}>0$ and b is a positive constant, the general solution of Eq. (14) is written as ${\phi }_{A\ }\left(x\right)\propto e^{\alpha x}$   where
\begin{equation}
{\alpha }^{\pm }=\frac{2m_0t_0a_0\tau \xi }{{\hbar }^2\beta }\pm \frac{2m_0}{{\hbar }^2\beta }{\left(t^2_0a^2_0+{\hbar }^2\beta (\Delta +\lambda \tau s)/2m_0\right)}^{1/2}, 
\end{equation} 
or
\begin{equation}
{\phi }_{A\ }\left(x\right)=e^{\left(2t_0a_0\tau \xi m_{0/{\hbar }^2\beta }\right)x}+\left[Ae^{\eta x}+Be^{-\eta x}\right], 
\end{equation} 
where, $A$ and $B$ are integration constants and
\begin{equation}
\eta =  \frac{2m_0}{{\hbar }^2\beta }{\left(t^2_0a^2_0+{\hbar }^2\beta (\Delta +\lambda \tau s)/2m_0\right)}^{1/2}. 
\end{equation} 
Now if we demand that the wave function is localized at the edges of the nanoribbon which are placed at $x=\pm a$ and ${\phi }_{A}(x=0)=0$, then 
\begin{equation}
{\phi }_{A\ }\left(x\right)=Ne^{\left(2t_0a_0\tau \xi m_{0}/({\hbar }^2\beta) \right)x}{\mathrm{sinh} \eta x\ }\  , 
\end{equation} 
where $N$  is a normalization constant. The sign of ${\xi }/{\beta }$ is determined to make the wave function finite in the limit $\left|x\right|\to \infty $.  If we consider $\xi >0\ $then $\beta <0$.
First, we define the general solution without $\xi $ factor. It means that only for $\beta <0$ a physical wave function exists which is localized at the edges of the nanoribbon. It has been shown that $\beta =-1.54$ \cite{R22}.
Using Eq. (11), since always $\beta <0$, then
\begin{equation}
C^{\tau }_s=\frac{\tau s}{2}+\frac{\tau s}{2}(sgn{\mathrm{(}\mathrm{\Delta }}^{\mathrm{\tau}}_{\mathrm{s}}\mathrm{)}).                                                                                    
\end{equation} 
Thus, if we take the spin as a good quantum number and consider one valley, the MoS${}_{2}$ monolayer has a non-trivial topology. This non-trivial topology results in crossing edge modes, which has been seen before \cite{R22}. As Eq. (23) shows $C^K_s=-C^{K^{\prime}}_s$ and as a consequence, the total Chern number is zero which is consistent with the time reversal symmetry \cite{R22}. But, $C_v\neq 0$ for each spin (see Eq. (16c)), and the total Chern number is zero.

\begin{figure*}[]
\captionsetup{singlelinecheck = false, justification=raggedright}
\includegraphics[width=0.5\linewidth]{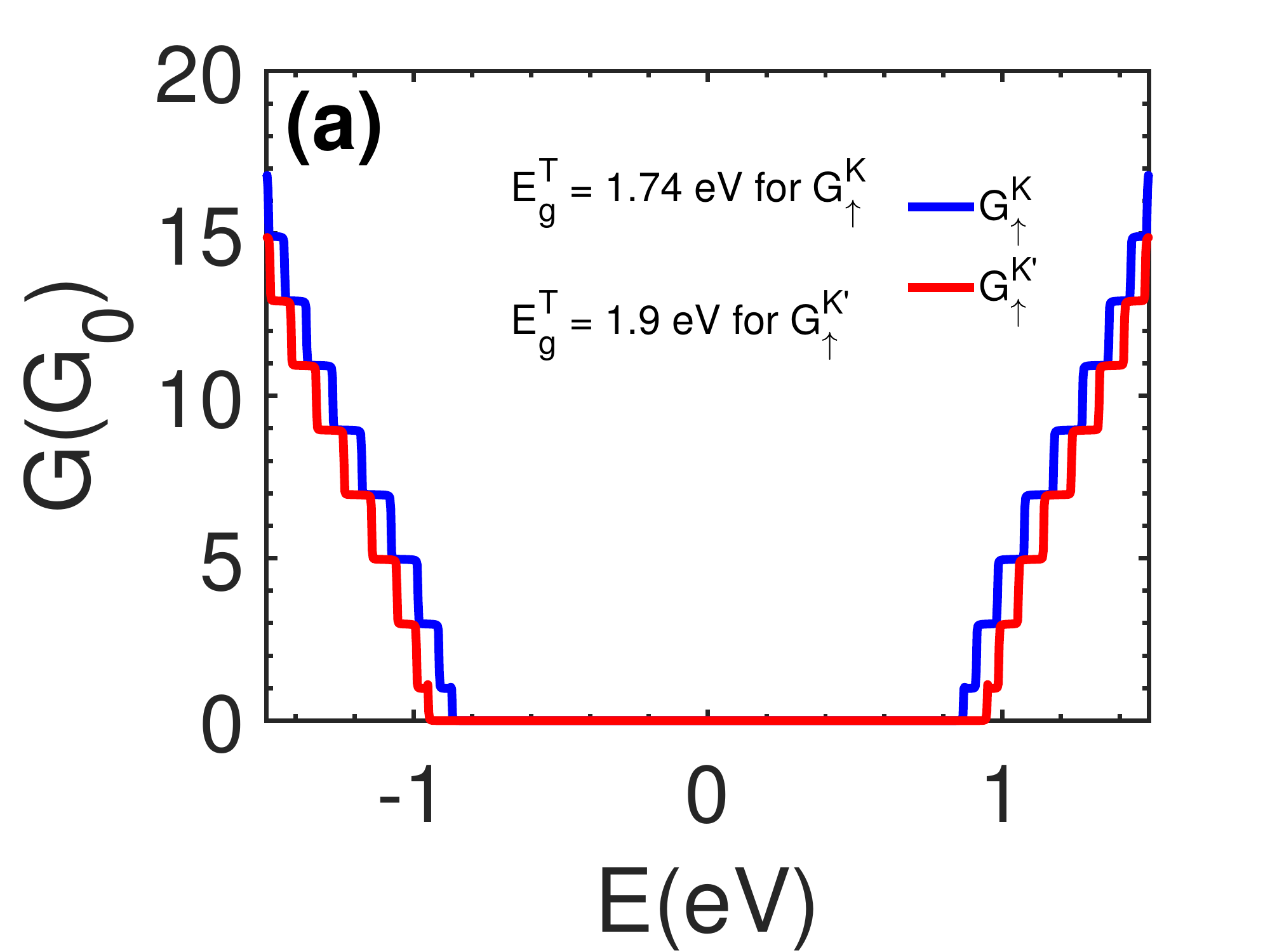}
\hspace{20pt}
\includegraphics[width=0.5\linewidth]{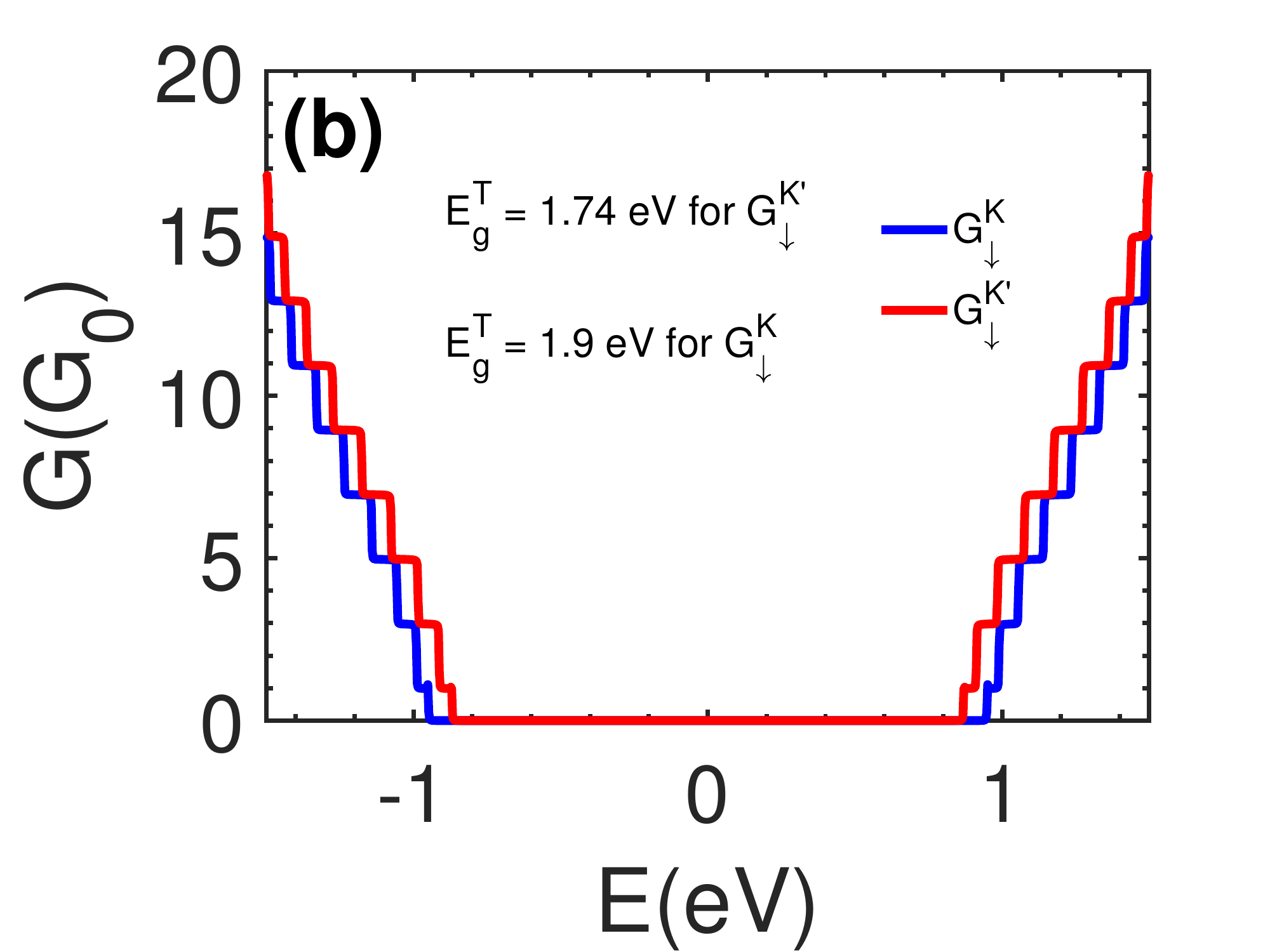}
\caption{(Color online) Quantum conductance of a monolayer MoS${}_{2}$ zigzag nanoribbon, for (a) spin up and (b) spin down. Here, $\Delta M=0$,  $\Delta=1.82$ eV and $\lambda =-0.08$ \cite{R22}. ($G_{0}=e^2/h$)}
\end{figure*}
\begin{figure*}[]
\captionsetup{singlelinecheck = false, justification=raggedright}
\includegraphics[width=0.5\linewidth]{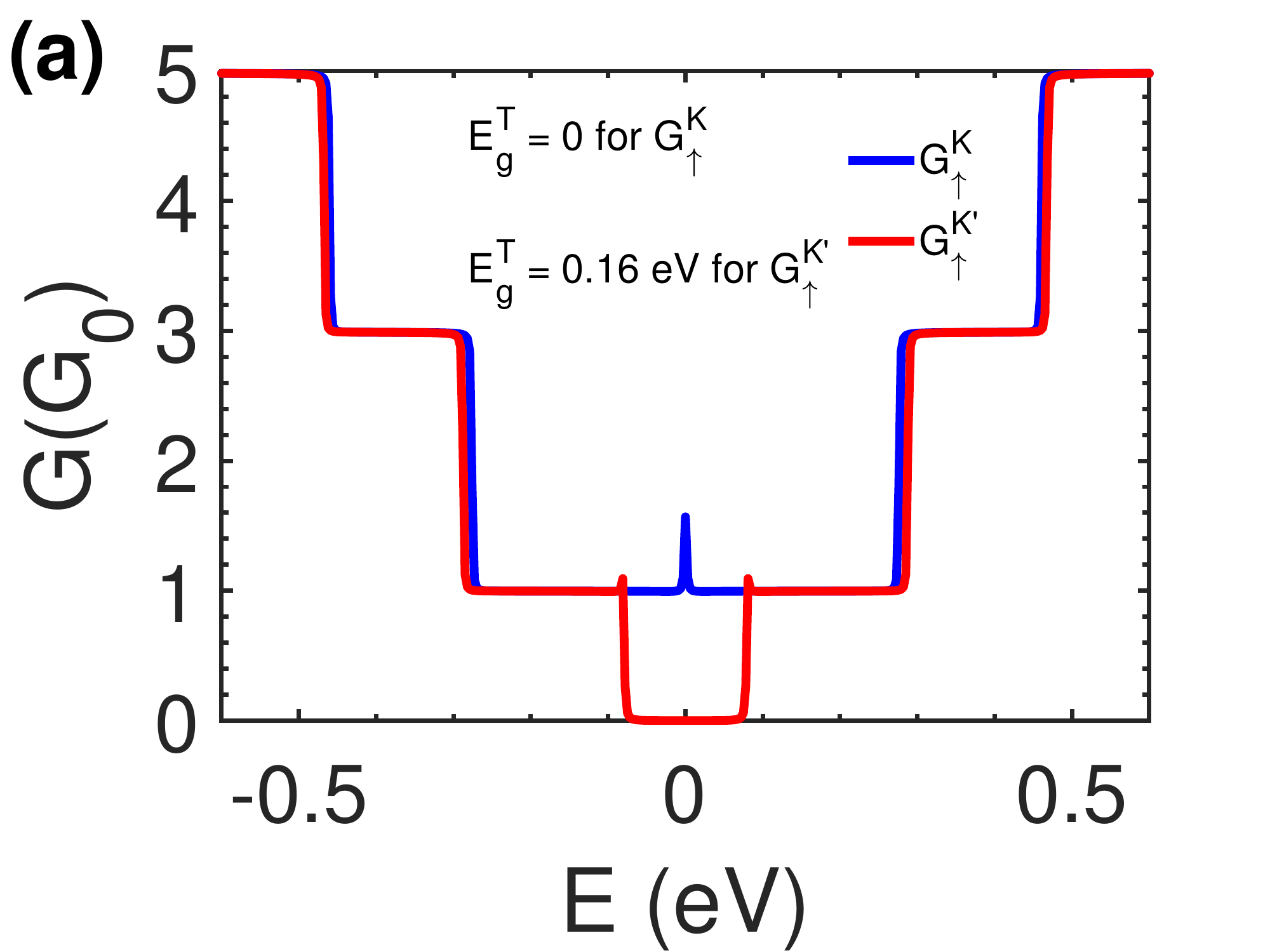}
\hspace{20pt}
\includegraphics[width=0.5\linewidth]{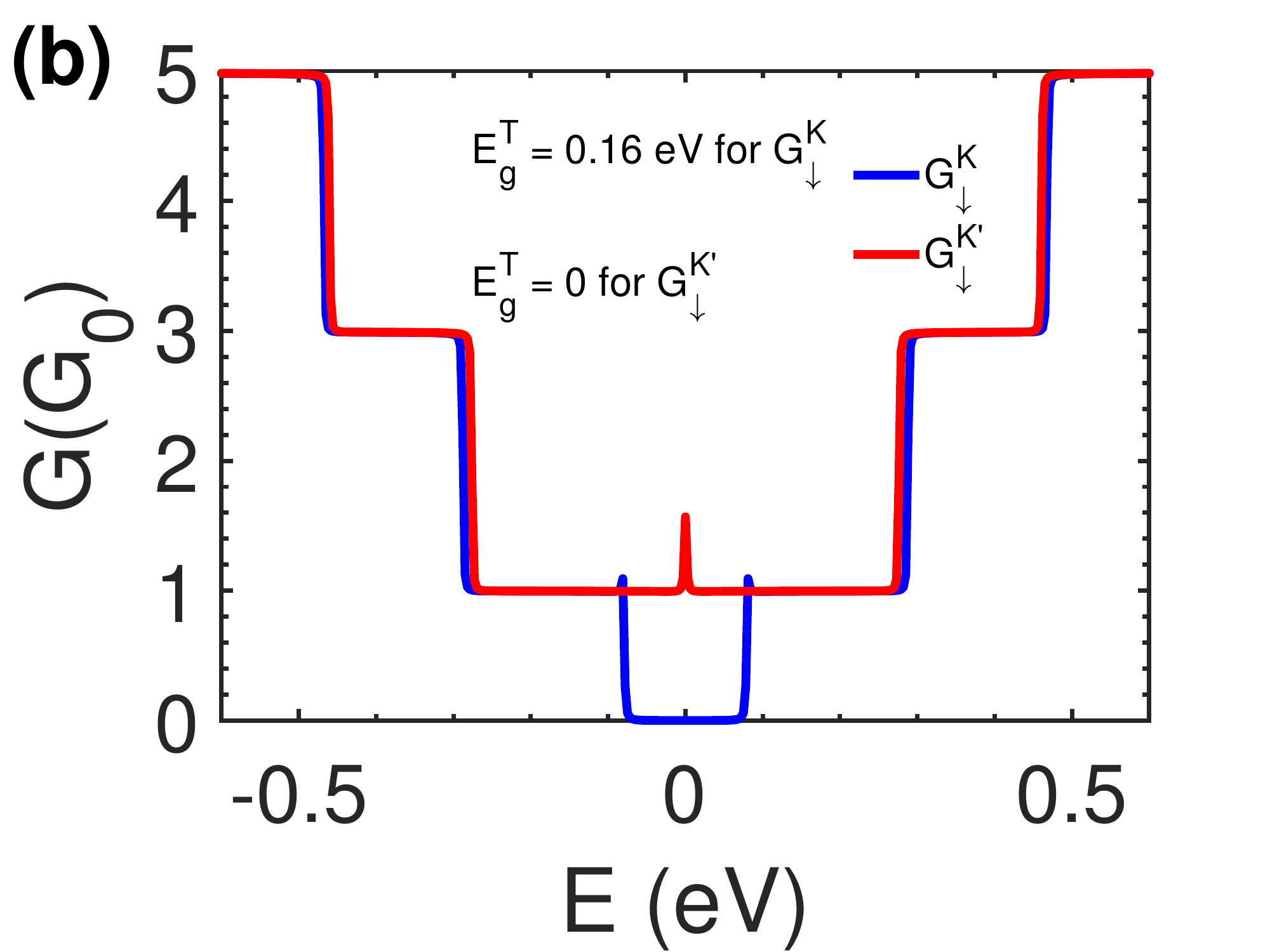}
\caption{(Color online) Quantum conductance of a monolayer MoS${}_{2}$ zigzag nanoribbon, for (a) spin up and (b) spin down. Here, $\Delta=1.82$ eV and $\lambda=-0.08$ \cite{R22}. It is assumed that a total field, $-\Delta M=(\Delta+\lambda)/2$, is applied to the nanoribbon. ($G_{0}=e^2/h$)}
\end{figure*}

\section{Topological quantum transistor}
Fig. 2(b) shows the different phases in the phase diagram for $\beta =0$. There are specific boundaries between regions such that at each boundary ${\Delta }^{\tau }_s=0$. Inside each boundary ${\Delta }^{\tau }_s\neq 0$ and consequently it is expected that there is a transmission gap in the quantum conductance curves when we cross each region. During a phase transition which is induced by applying a specific set of$\ (V,\ \Delta M)$ parameters, the variation of ${\Delta }^{\tau }_s$ is seen as  ${\Delta }^{\tau }_s\neq 0\to {\Delta }^{\tau }_s=0\to {\Delta }^{\tau }_s\neq 0$. However, the condition ${\Delta }^{\tau }_s\neq 0$, which means the conductance is zero at the transmission gap, which can be considered as an ``Off" state and the condition ${\Delta }^{\tau }_s=0$, which means the conductance is non-zero at the transmission gap, which is like an ``On" state. Therefore, we are able to design a quantum topological field effect transistor by using a monolayer MoS${}_{2}$ and applying a specific set of $(V,\ \Delta M)$ values (similar to published results in silicene \cite{R15,R16,R17}). 
\\Now, we consider a zigzag monolayer of MoS${}_{2}$ nanoribbon and assume a total field $-\Delta M=(\Delta +\lambda )/2$ is applied to the nanoribbon. We can calculate the quantum conductance of the nanoribbon by using the tight-binding non-equilibrium Greens function (NEGF) method [33, 35]. Fig. 3 (for $\Delta M=0$) and 4 (for $-\Delta M=(\Delta +\lambda )/2$) show the quantum conductance as function of the fermi energy. As Fig. 3(a) shows, there is a transmission gap equal to 1.74 (1.9) eV for spin up electrons from $K$ ($K^\prime$)-valley and their quantum conductance is higher (lower) than the quantum conductance of spin up electrons from $K^\prime$ ($K$)-valley. Fig. 3(b) shows the tranmission gap and quantum conductance of spin down electrons from $K$ and $K^\prime$-valleys. Here, the conductance of spin down electrons from $K^\prime$-valley is higher than the quantum conductance of spin up electrons from $K$-valley.   
 As Fig. 4(a) (Fig. 4(b)) shows, the transmission gap is closed by applying the exchanged field $\Delta M$ for the spin up (down) electrons from the $K$($K^\prime$)-valley. Therefore, spin up electrons from $K$-valley and the spin down electrons from the $K^{\prime}$-valley take part in the quantum conductance because the transmission gap, $E^T_g$, is zero for these kinds of spins. Since both kinds of spins take part in the conductance, we deal with helical edge states.
When $\beta \neq 0$, we cannot use the NEGF method because the quadratic ($q^2$) diagonal term is present in the two-band Hamiltonian model. Previously, some of us showed that the quantum conductance and the related edge states can be calculated by using a six-band Hamiltonian model \cite{R29} when $\beta \neq 0$. They have shown that how one can close the transmission gap by applying an external electric field and an exchanged field \cite{R29}.

\section{Conclusion}
We obtained topological phase transitions in a monolayer of MoS${}_{2}$ within a two-band Hamiltonian model. By neglecting the electron-hole asymmetry and quadratic ($q^2$) diagonal terms in the spin-valley Dirac mass equation, we showed that the phase diagram includes QAH, QSH, and SQAH regions such that the topological Kirchhoff law is satisfied in the plane. In this case, we found that the wave function is localized inside (not at the edges) of a nanoribbon when an external potential $V\left(x\right)\propto x$ is applied along the width of the nanoribbon. \\By adding the quadratic ($q^2$) diagonal terms, we found that MoS${}_{2}$ has a non-trivial topology if one valley is considered and spin is a good quantum number. In this case, the wave function is localized at the edges of the nanoribbon.
\\We considered a zigzag monolayer of MoS${}_{2}$ and studied the quantum transport by using the NEGF method. We showed that the spin-valley Dirac mass term (${\Delta }^{\tau }_s)$ could be zero by applying an external potential and an exchange field such that we deal with helical edge states in the nanoribbon. The ${\Delta }^{\tau }_s\neq 0$ (${\Delta }^{\tau }_s=0$) case has been assigned to ``Off (On)" state of a field effect transistor. Therefore, the device can act as a field effect topological quantum transistor.  \\ \\

\nocite{*}
\bibliographystyle{iopart-num}
\bibliography{article}

\end{document}